\documentclass[11pt]{article}
\usepackage{moriond,epsfig}
\usepackage{physics}

\def\Journal#1#2#3#4{{#1} {\bf #2}, #3 (#4)}

\def\PLB{{\em Phys. Lett.}  B}
\def\PRL{\em Phys. Rev. Lett.}
\def\PRD{{\em Phys. Rev.} D}

\def\sst{\scriptscriptstyle}

\def\ra{\rightarrow}

\def\be{\begin{equation}}
\def\ee{\end{equation}}
\def\bea{\begin{eqnarray}}
\def\eea{\end{eqnarray}}

\newcommand{\q}{ Q^2 }
\newcommand{\mgg}{ W_{\gamma \gamma }}
\newcommand{\gamgam}{\gamma \gamma^*}

\begin{document}
\vspace*{4cm}
\title{EXCLUSIVE {\boldmath$\rho\rho$} PRODUCTION IN TAGGED
{\boldmath$\gamma\gamma$} INTERACTIONS AT LEP}

\author{ I. VOROBIEV
\footnote{\emph{Present address:} CERN, CH-1211 Geneva 23,
 Switzerland; e-mail: Igor.Vorobiev@cern.ch}
(on behalf of the L3 Collaboration) }

\address{Carnegie Mellon University, Pittsburgh, PA 15213}

\maketitle\abstracts{
Exclusive $\rho\rho$ production in two-photon collisions between a
quasi-real, $\gamma$, and a virtual, $\gamma^*$, photons is studied at LEP
at centre-of-mass energies \mbox{$89 \GeV \le \sqrt{s} \le 209 \GeV{}$}
with a total integrated luminosity of $854.7~\pb$.  The cross sections of the
$\, \gamma \gamma^* \rightarrow \rho\rho \,$ processes are
determined as a function of the photon virtuality, $\q$, and the
two-photon centre-of-mass energy, $\mgg$, in the kinematic region:
$0.2 \GeV^2  \le  \q  \le 30\GeV^2$ and
$1.1 \GeV \le \mgg \le 3 \GeV$.
}

\section{Introduction}
     Recently a QCD model was proposed \cite{mdiehl} for calculating the
cross section of exclusive meson pair production in gamma-gamma interactions
with one highly virtual photon having $\q \gg \mgg$. In this model, the
exclusive process is factorisable into a perturbative, calculable,
short distance scattering $\rm \gamma\gamma^*\rightarrow q\bar{q}$ or 
$\rm \gamma\gamma^*\rightarrow gg$ and non-perturbative matrix elements,
which are called generalized distribution amplitudes, describing the
transition of the two partons into hadron pairs.

The L3 Collaboration performed a series of measurements of neutral and
charged $\rho$-meson pair production in processes
$$e^+ e^- \rightarrow e^\pm {e^\mp}_{tag} \: \gamma \gamma^* \rightarrow
e^\pm {e^\mp}_{tag} \: \rho \rho \, ,$$
where one electron/positron is tagged either by Luminosity Monitor (LUMI)
or Very Small Angle Tagger (VSAT) ($\rho^0\rho^0$ with LUMI - 
Ref.\cite{l3-269}, $\rho^+\rho^-$ with LUMI - Ref.\cite{l3-287},
$\rho^0\rho^0$ with VSAT - Ref.\cite{l3-292}, $\rho^+\rho^-$ with VSAT
- Ref.\cite{l3-299}). LUMI covers $\q$ range $1.2 - 8.5 \GeV^2$ at LEPI,
$8.8 - 30 \GeV^2$ at LEPII, and VSAT (available only at LEPII) -
$0.2-0.85 \GeV^2$.
\begin{figure}[t]
\vskip 2.5cm
\begin{tabular}{cc}
\hspace{2cm} \epsfig{figure=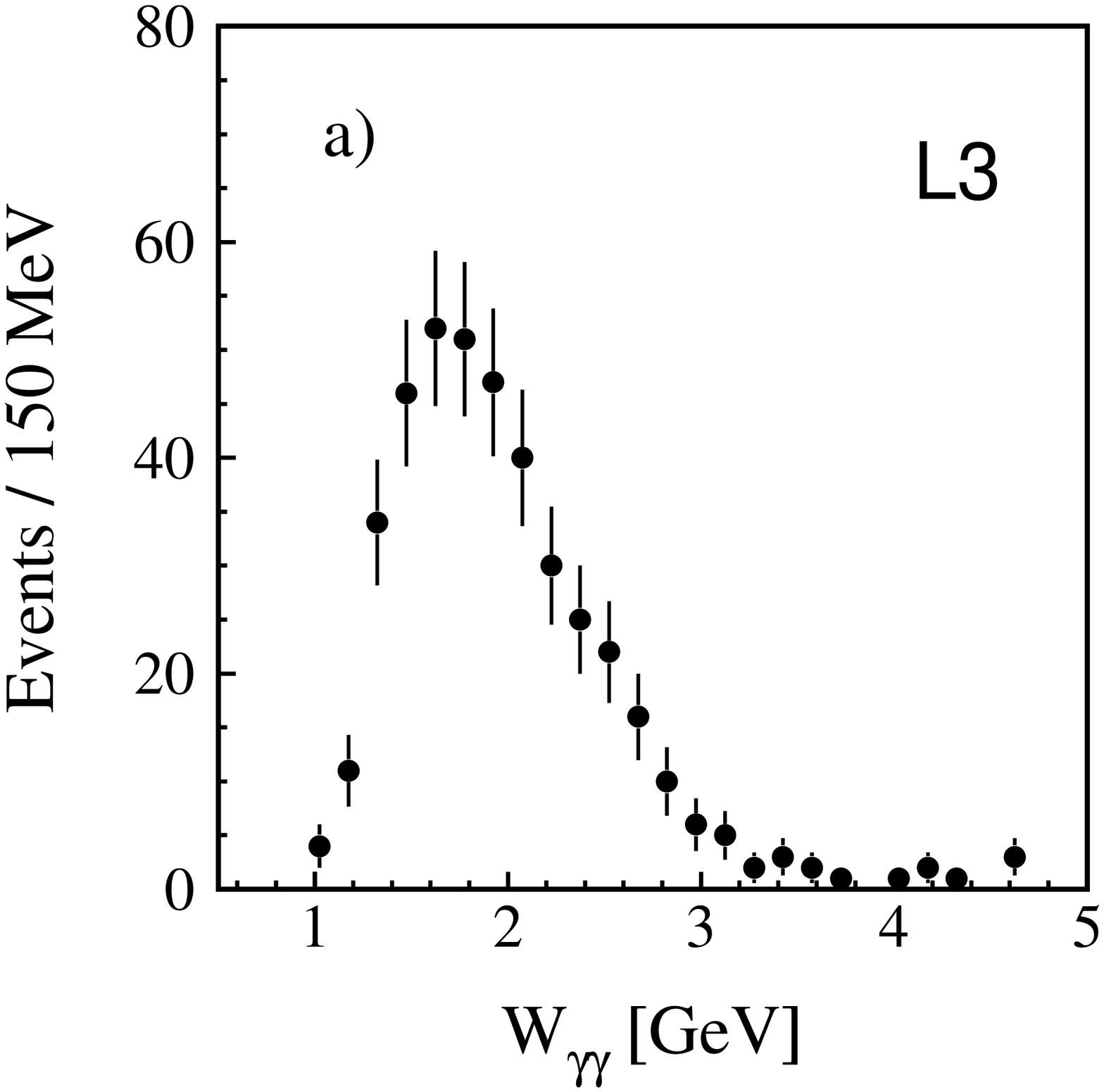,height=2.0in}
&
\epsfig{figure=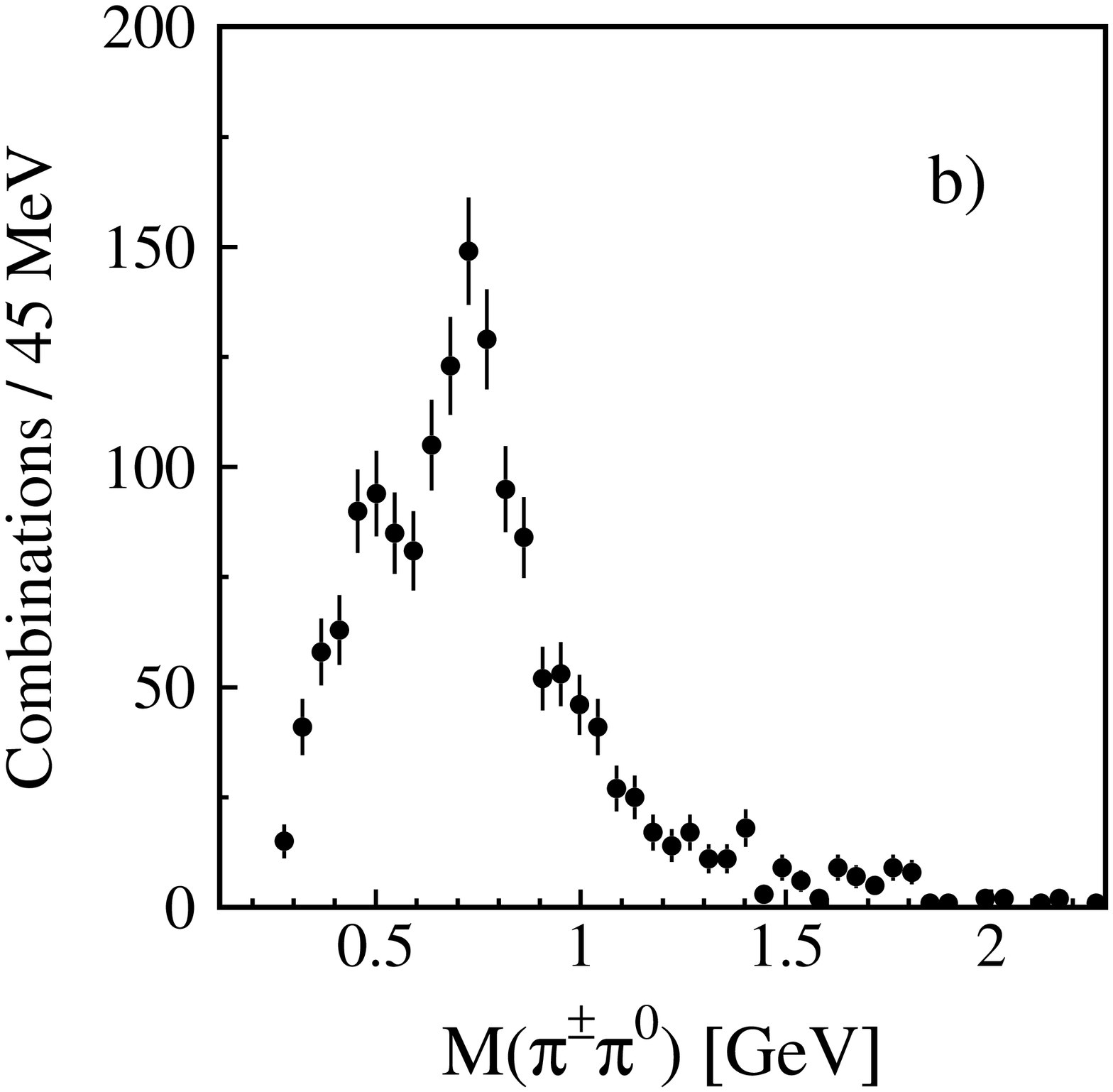,height=2.0in} \\
\hspace{2cm} \epsfig{figure=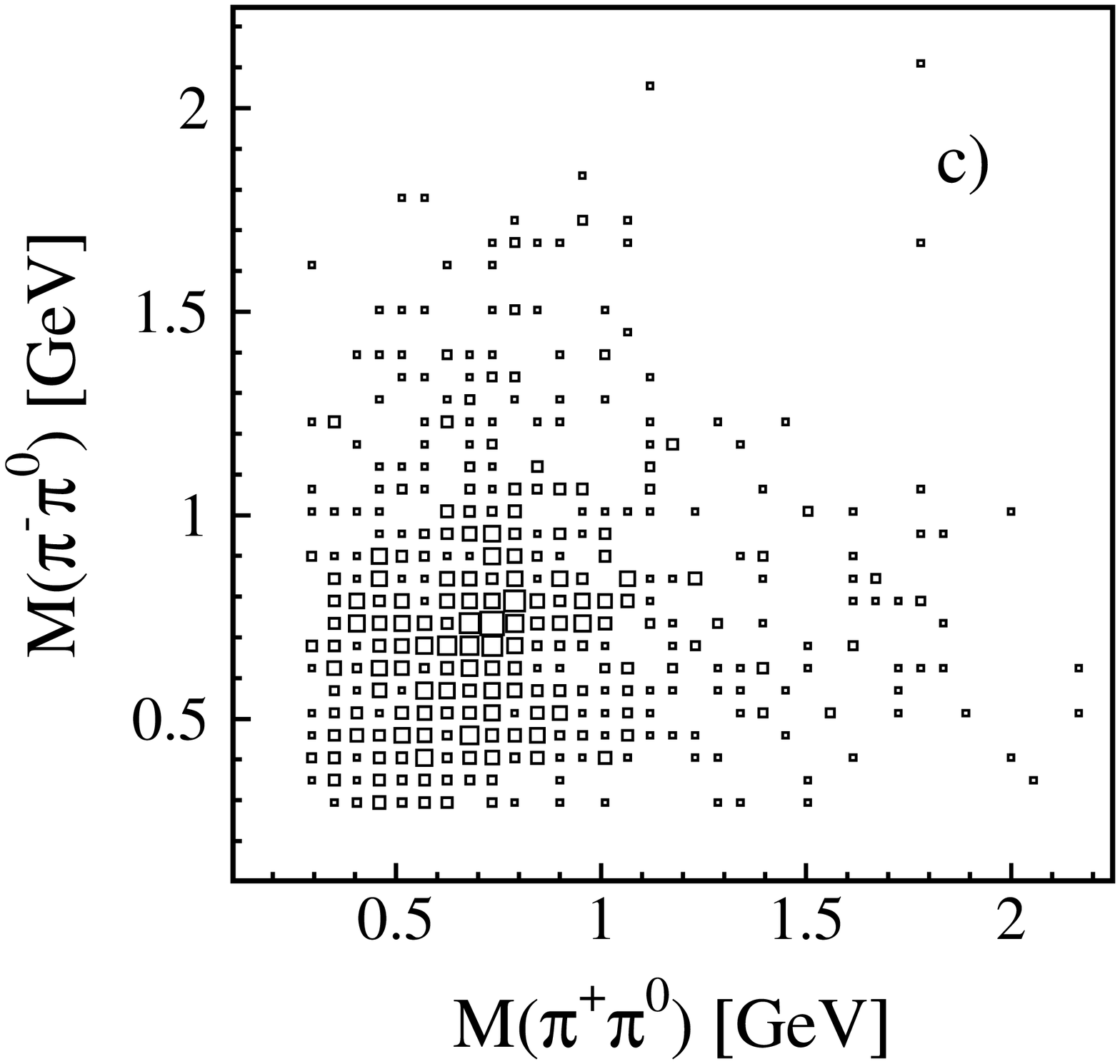,height=2.0in}
&
\epsfig{figure=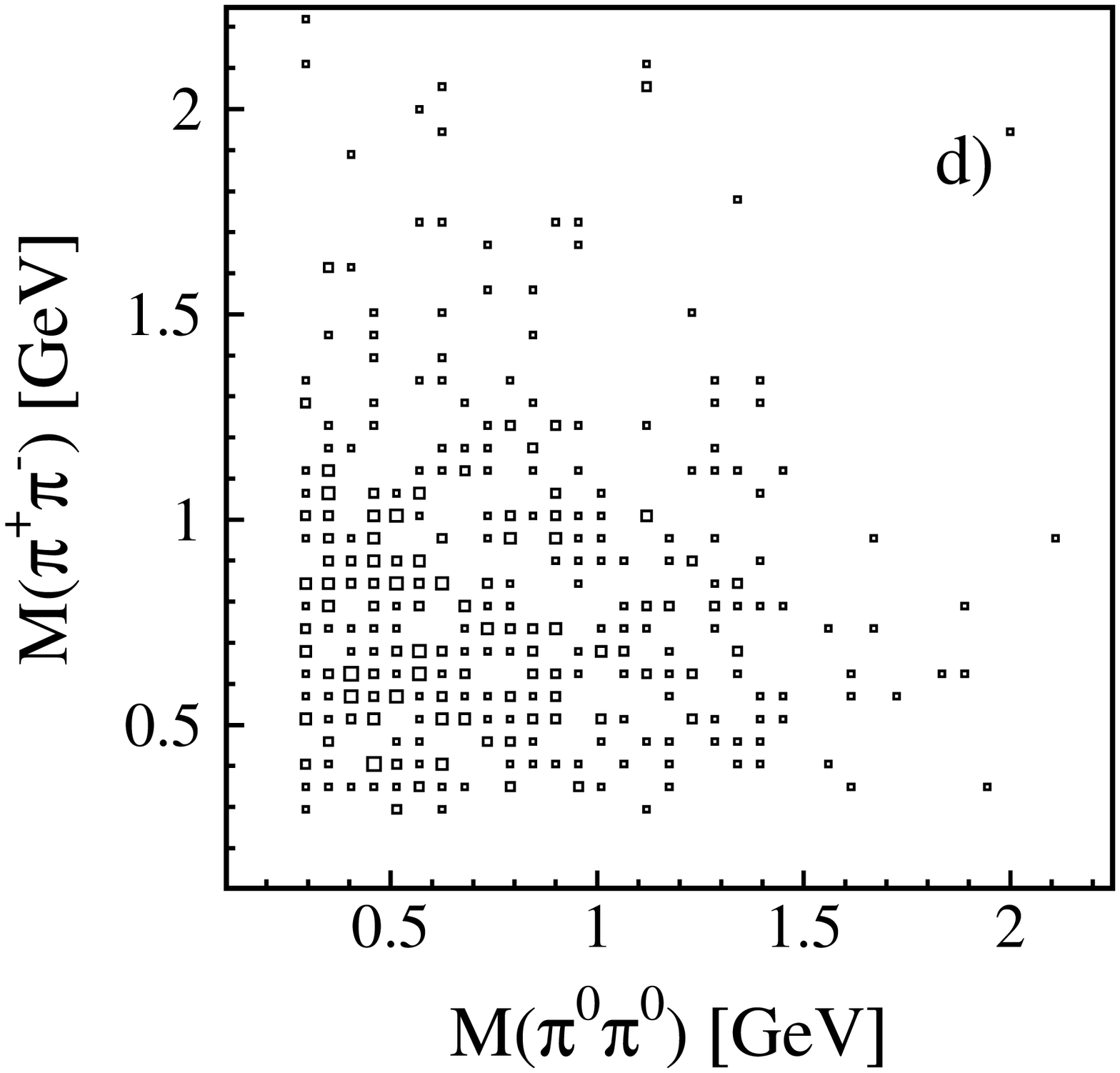,height=2.0in}
\end{tabular}
\caption{Mass distributions for the selected $\pi^+\pi^-\pi^0\pi^0$ events
with VSAT tag:
a) the four-pion system mass, $\mgg$;
b) the $\pi^\pm \pi^0$ combinations (four entries per event);
c) correlation between  the $\pi^-\pi^0$ and  $\pi^+\pi^0$ pairs
 (two entries per event) and
d) correlation between  the $\pi^+\pi^-$ and  $\pi^0\pi^0$ pairs.
\label{fig:sel}}
\end{figure}

\section{Measurements}
Measurements with LUMI were done at LEPI ($\sqrt{s}$ around 91$\GeV$)
with integrated luminosity $L = 148.7 pb^{-1}$ and at LEPII
(\mbox{$161 \GeV \le \sqrt{s} \le 209 \GeV{}$}) with $L = 706.0 pb^{-1}$, 
measurements with VSAT - at LEPII (\mbox{$183 \GeV \le \sqrt{s}
\le 209 \GeV{}$}) with $L = 684.8 pb^{-1}$. Underlying selected reactions are
\begin{eqnarray}
\label{rea1}
e^+ e^- \rightarrow e^\pm {e^\mp}_{tag} \: \pi^+\pi^-\pi^+\pi^- \\
\label{rea2}
e^+ e^- \rightarrow e^\pm {e^\mp}_{tag} \: \pi^+\pi^-\pi^0\pi^0
\end{eqnarray}
934 events of reaction (\ref{rea1}) were selected in LUMI sample and 1938 -
in VSAT one. For reaction (\ref{rea2}) - 343 and 414 events respectively.
Example of distributions for selected events (for reaction (\ref{rea2})
with VSAT) is shown in Figure~\ref{fig:sel}. Mass of charged pion pairs
(Figure~\ref{fig:sel}b) shows a strong signal of $\rho^\pm$  production, and
two-dimensional distribution of such masses (Figure~\ref{fig:sel}c) presents
concentration of events in region of $\rho^+\rho^-$ signal. No structure
is observed in the correlation plot of $\pi^+\pi^-$ and $\pi^0\pi^0$ mass
combinations (Figure~\ref{fig:sel}d). For further analysis and cross section
determination events in the region
$1.1 \GeV  \le \mgg \le 3 \GeV$ were left (Figure~\ref{fig:sel}a).

In order to determine the differential $\rho\rho$ production rates, a
maximum likelihood fit was performed in intervals of $\q$ and $\mgg$
using a box method.
Data were fit to a sum of non-interfering contributions from the processes,
generated by Monte Carlo in simple model of isotropic production and
phase space decay. For data sample with LUMI the processes used in fit were:
for the reaction (\ref{rea1}): 
$\gamma \gamma^* \rightarrow \rho^0 \rho^0$,
$\gamma \gamma^* \rightarrow \rho^0 \pi^+ \pi^-$,
$\gamma \gamma^* \rightarrow \pi^+ \pi^- \pi^+ \pi^-$ (non resonant);
for the reaction (\ref{rea2}): 
$\gamma \gamma^* \rightarrow \rho^+ \rho^-$,
$\gamma \gamma^* \rightarrow \rho^\pm \pi^0 \pi^0$,
$\gamma \gamma^* \rightarrow \pi^+ \pi^- \pi^0 \pi^0$ (non resonant).
For data sample with VSAT it was found that a fit is improved if to include
additional channels - for the reaction (\ref{rea1}):
$\gamma \gamma^* \rightarrow f_2(1270) \rho^0$,
$\gamma \gamma^* \rightarrow f_2(1270) \pi^+ \pi^-$
and for the reaction (\ref{rea2}):
$\gamma \gamma^* \rightarrow {a^\pm}_2(1320) \pi^0$ (we do not determine
the cross sections of these processes).
The inputs to the fit were the six possible two-pion mass combinations
in an event. Fit provides a good description of all mass and angular
distributions. From numbers of $\rho\rho$ events, obtained by fit in
intervals of $\q$ and $\mgg$ and corrected by efficiencies and background,
cross sections of $\rho\rho$ pair production were calculated.

\begin{figure}[t]
\vskip 2.5cm
\begin{tabular}{cc}
\psfig{figure=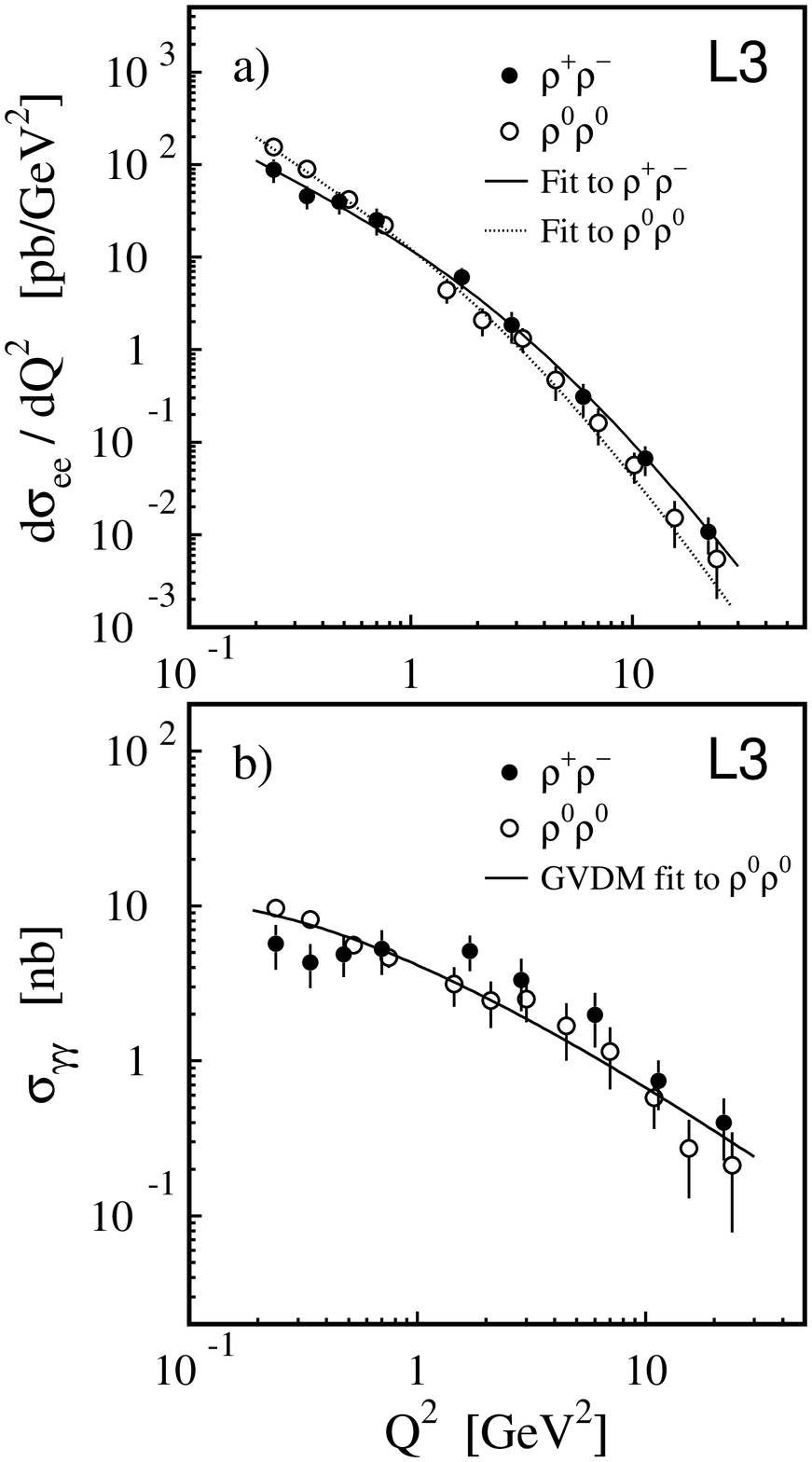,height=4.5in}
&
\epsfig{figure=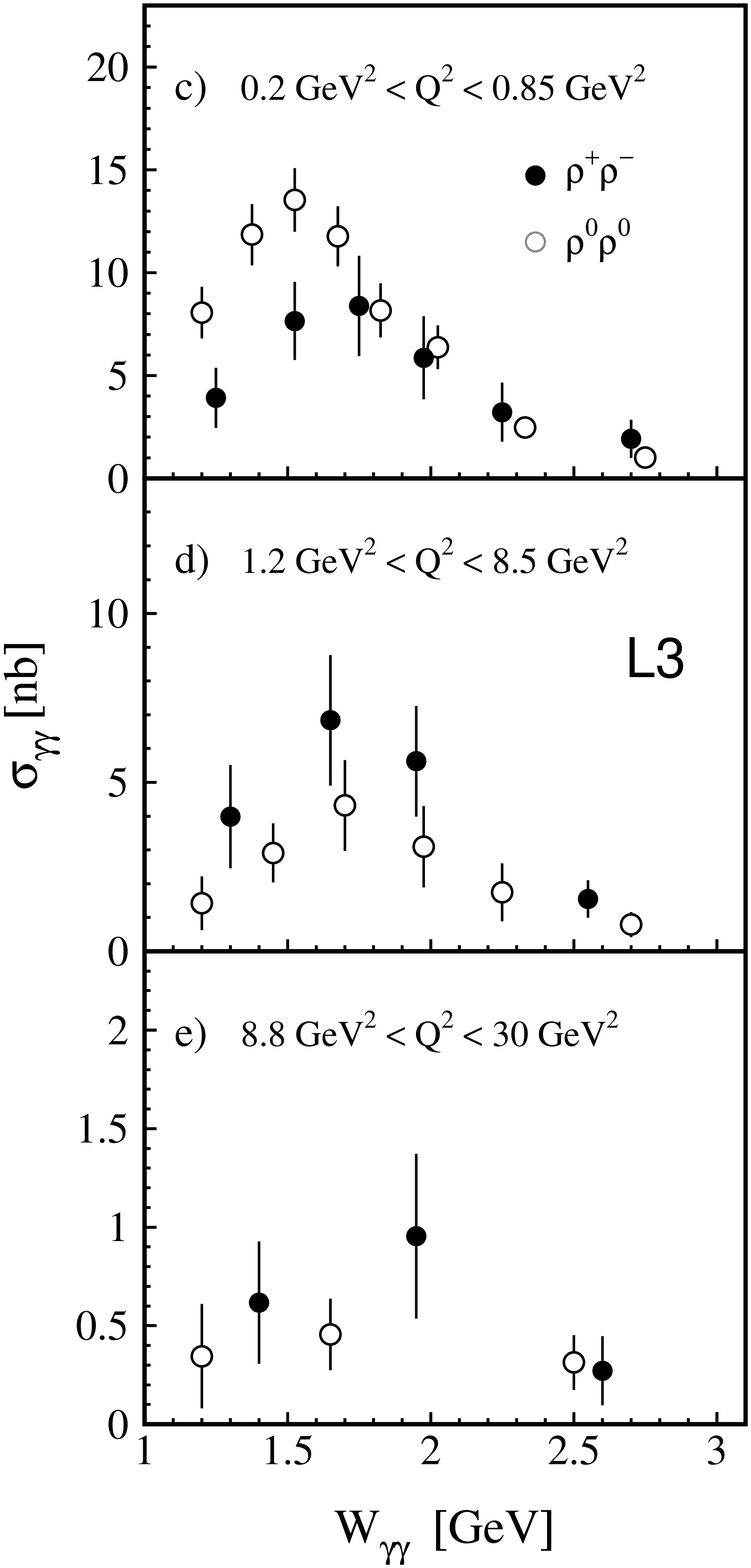,height=4.3in}
\end{tabular}
\caption{The $\rho\rho$ production cross section as a function of $\q$:
a) differential cross section of the process
$e^+e^-  \to e^+e^- \rho\rho$ and
b) cross section of the process $\gamma\gamma^* \to \rho\rho$.
c)d)e) Cross section of the process $\gamma\gamma^* \to \rho\rho$
as a function of  $\mgg$ in three $\q$ intervals.
\label{fig:res}}
\end{figure}

\section{Results}
The cross section dependence on $\q$ and $\mgg$ for both $\rho^0\rho^0$
and $\rho^+\rho^-$ is shown on Figure~\ref{fig:res}. $\q$ dependence of
differential cross section can be well described by the approximate formula
$d \sigma_{\mathrm{ee}} / d \q \sim 1 / Q^n(\q + < \mgg >^2)^2 \,$, following
from QCD-based calculations \cite{diehl2}, where $n$ is expected to be equal
2. In the range $\q > 1.2 \GeV^2$ fit gives values $n=2.4 \pm 0.3$ for
$\rho^0 \rho^0$ and $n=2.5 \pm 0.4$ for $\rho^+ \rho^-$, which are compatible
with the expected value 2. $\rho^0\rho^0$ production at $\q > 1.2 \GeV^2$
was analyzed in paper \cite{anikin}. $\q$ dependence of the differential
cross section is perfectly reproduced by formulas with three phenomenological
parameters, one of which, $C_1 = 1.20 \pm 0.23 \GeV^2$, gives normalization
of $\rho\rho$ generalized distribution amplitudes.

Ratio of cross sections $\sigma_{e^+ e^- \rightarrow
e^+ e^- \rho^+\rho^-}/\sigma_{e^+ e^- \rightarrow e^+ e^- \rho^0\rho^0}$
(in the kinematic region $1.2 \GeV^2  \le  \q  \le 8.5 \GeV^2$, $1.1 \GeV
\le \mgg \le 2.1 \GeV$) is $1.8 \pm 0.5$, which is compatible with the
factor 2, expected for an isospin $I=0$ state. Contrary to this
measurement in high $\q$ region, in previous measurements done for
$\q \sim 0$ it was found that $\rho^0\rho^0$ cross section
is few times higher than $\rho^+\rho^-$ one. And really, when we
go to smaller $\q$, the relative magnitude of $\rho^+\rho^-$ and
$\rho^0\rho^0$ production changes in the vicinity of 
$\q \approx 1 \GeV^2$, suggesting different $\rho$-pair production
mechanisms at low and high $\q$.
In the whole $\q$ range fit to approximate formula gives
$n=2.3 \pm 0.15$ for $\rho^+ \rho^-$ and $n=2.9 \pm 0.14$ for
$\rho^0 \rho^0$ (fit is shown in Figure~\ref{fig:res}a by lines). 

The measured cross section of the process $ \gamgam \to \rho \rho$ as a
function of $\q$ is shown in Figure~\ref{fig:res}b. It is well described by
GVDM model both for $\rho^+\rho^-$ and $\rho^0\rho^0$ at $\q > 1 GeV^2$,
and for $\rho^0\rho^0$ - in the whole $\q$ range. The fit for $\rho^0\rho^0$
in the whole $\q$ range finds cross section $13.6 \pm 0.7$ nb at $\q = 0$.
The change of relative magnitude of $\rho^+\rho^-$ and $\rho^0\rho^0$
processes is clearly visible, when comparing low- and high-$\q$ regions.
A $\rho$-pole description (is not shown on the picture) is excluded for both
$\rho^0\rho^0$ and $\rho^+\rho^-$ data.

The cross sections of the processes $ \gamgam \to \rho \rho$ are plotted in
Figure~\ref{fig:res}c-e as a function of $\mgg$. For $\mgg  \le 2.1 \GeV$
on Figure~\ref{fig:res}c there is a clear enhancement of $\rho^0\rho^0$
production relative to $\rho^+\rho^-$. Ratio $\sigma_{e^+ e^- \rightarrow
e^+ e^- \rho^+\rho^-}/\sigma_{e^+ e^- \rightarrow e^+ e^- \rho^0\rho^0}
= 0.6 \pm 0.1$ in this region.

\section{Conclusions}
This is the first measurement of tagged exclusive $\rho^0\rho^0$ production
and of $\rho^+\rho^-$ production at high $\q$.
The measurements allow to follow the evolution of cross sections over a
$\q$-region of two orders of magnitude. A QCD-based form is found to provide
a good parametrization in the entire $\q$ interval, where the differential
cross sections show a monotonic decrease of more than four orders of magnitude
(for the $\mgg$ range $1.1 \GeV \le \mgg \le 3 \GeV$). The $\q$ dependence
of the process $\gamma \gamma^* \to \rho^0 \rho^0$ is well reproduced by
parametrization based on GVDM model over the entire $\q$-region. But
$\rho^+\rho^-$ data cannot be satisfactory described by such a parametrization
in the whole $\q$-region, only at $\q > 1 \GeV^2$. A $\rho$-pole
description is excluded for both $\rho^0\rho^0$ and $\rho^+\rho^-$ data.
The relative magnitude of $\rho^+\rho^-$ and $\rho^0\rho^0$ production
changes in the vicinity of $\q \approx 1 \GeV^2$, suggesting different
$\rho$-pair production mechanisms at low and high $\q$. These data lay
a new experimental grounds for obtaining information about QCD going from
non-perturbative to the perturbative regime.

\end{document}